\documentclass[aip,reprint,jcp,amsmath,graphicx]{revtex4-1}

\usepackage{graphicx}
\usepackage{dcolumn}
\usepackage{bm}
\usepackage{color}

\draft 

\begin{document}

\title{Nonlinear polarization evolution using time-dependent density functional theory}

\author{Mitsuharu Uemoto}
\email[]{uemoto@ccs.tsukuba.ac.jp.}
\affiliation{Center for Computational Sciences, University of Tsukuba, 1-1-1 Tennodai, Tsukuba, Ibaraki, Japan.}

\author{Yuki Kuwabara}
\affiliation{Graduate School of Pure and Applied Sciences, University of Tsukuba, 1-1-1 Tennodai, Tsukuba, Ibaraki, Japan.}

\author{Shunsuke A. Sato}
\affiliation{Max Planck Institute for the Structure and Dynamics of Matter, Luruper Chaussee 149, D-22761 Hamburg, Germany.}

\author{Kazuhiro Yabana}
\affiliation{Graduate School of Pure and Applied Sciences, University of Tsukuba, 1-1-1 Tennodai, Tsukuba, Ibaraki, Japan.}

\date{\today}

\begin{abstract}
  We propose a theoretical and computational approach to investigate temporal behavior of a nonlinear polarization
  in perturbative regime induced by an intense and ultrashort pulsed electric field.
  First-principles time-dependent density functional theory is employed to describe the electron dynamics.
  Temporal evolution of third-order nonlinear polarization is extracted from a few calculations of
  electron dynamics induced by pulsed electric fields with the same time profile but different amplitudes.
  We discuss characteristic features of the nonlinear polarization evolution as well as an extraction of
  nonlinear susceptibilities and time delays by fitting the polarization.
  We also carry out a decomposition of temporal and spatial changes of the electron density in power series
  with respect to the field amplitude. It helps to get insight into the origin of the nonlinear polarization in atomic scale.
\end{abstract}


\maketitle 

\section{Introduction}
\label{sec:intro}

Nonlinear polarization is a fundamental quantity that characterizes the interaction
of a high intensity light with bulk materials \cite{boyd2003nonlinear, butcher1991elements}.
In early studies, measurements have been carried out for nonlinear susceptibilities
in frequency domain, $\chi^{(n)}(\omega)$, using a sufficiently long pulsed light that
can be regarded as a continuous wave.
A number of applications utilizing nonlinear optical properties have been
developed including frequency conversion \cite{franken1961generation, chen2002second}, optical Kerr effect
\cite{spence199160}, and so on. There have been intensive attempts to find photonic
materials that have useful nonlinear optical properties \cite{chen2002second, yu20172d}.

Recently, owing to developments in ultrashort laser pulse technologies, it has become
possible to explore nonlinear polarization in time domain. Attosecond metrologies
\cite{krausz2009attosecond} have made it possible to explore  electron dynamics in crystalline solids
in time resolution less than a period of optical pulses \cite{lucchini2016attosecond, sommer2016attosecond, schultze2014attosecond, hofmann2015noninstantaneous}.
These studies aim, as an ultimate goal, to achieve information processing utilizing
ultrashort pulsed light. For such purposes, it is essentially important to establish experimental
and theoretical methods to explore the temporal evolution of the nonlinear polarization.

In theoretical side, there have been many efforts to describe, understand, and
predict nonlinear optical response of materials. In accord with the developments of
measurements and applications in frequency domain, theoretical efforts have been first devoted
to explore frequency-dependent nonlinear susceptibilities. Empirical formula that relate
nonlinear susceptibilities with the linear one introducing an anharmonicity coefficient have
been established \cite{miller1964optical, boling1978empirical}.
Quantum mechanical approaches of different levels of sophistication have been
developed for theoretical evaluations of frequency-dependent nonlinear susceptibilities.
Independent particle approximation (IPA) \cite{sipe1993nonlinear,aspnes1972energy,aversa1995nonlinear} is a well
established approach to calculate the susceptibilities using electron orbitals from the static electronic structure calculation.
Although it provides reasonable descriptions\cite{moss1987empirical}, it has been known that
there are difficulties in quantitative descriptions.
Significances of many-body effects such as the local field and excitonic effects
have been suggested \cite{leitsmann2005second, attaccalite2011real}.

In the last two decade, first-principles computational approaches based on
time-dependent density functional theory (TDDFT) \cite{runge1984density, yabana1996time} have been
developed  and applied to explore nonlinear polarizations.
Several computational methods have been developed to investigate
nonlinear polarizabilities in frequency domain: for molecules \cite{iwata2001real, andrade2007time}
and for solids \cite{dal1996density}. Time-domain methods solving the
time-dependent Kohn-Sham (TDKS) equation, the basic equation of TDDFT, in real time have
also been developed and applied to extract nonlinear polarizabilities in molecules \cite{takimoto2007real}
and in solids \cite{goncharov2013nonlinear, gruning2016dielectrics, attaccalite2013nonlinear}.

Recently, TDDFT have also been applied successfully to investigate temporal, ultrafast evolution
of nonlinear polarization in solids in femto- and attosecond time scale.
In \cite{schultze2014attosecond}, electron dynamics in crystalline silicon has been calculated and the results
are compared with measurements utilizing attosecond metrology.
In the previous work \cite{yabana2012time}, large-scale computations simultaneously solving the Maxwell's
equations for light electromagnetic fields and the TDKS equation for electron dynamics have
been carried out and compared with time-domain measurements to explore temporal evolution
of nonlinear polarization in dielectrics \cite{lucchini2016attosecond, sommer2016attosecond}.

Although TDDFT has been successful to describe nonlinear optical responses,
it has been recognized that inclusion of many-body correlation effects is not sufficient
in most exchange-correlation potentials employed in practice. To include long-range correlation
effects, for example, time-dependent density-polarization functional theory has been developed
and applied for nonlinear susceptibilities \cite{gruning2016dielectrics}.

In the present work, we propose a method based on TDDFT to investigate temporal evolution
of nonlinear polarization in perturbative regime induced by an intense and ultrashort pulsed light.
Carrying out a few calculations using pulsed electric fields with the same time profile but different amplitudes,
we numerically extract temporal evolution of nonlinear polarization in power series of the field
amplitude up to third order.
From the extracted nonlinear polarization components, we extract the coefficients of nonlinear
susceptibilities and the time-delay and compare them with measurements and previous calculations.
We also perform a power series expansion of the electron density changes.
Temporal and spatial distribution of the electron density change is expected to be useful to understand
the mechanism of the nonlinear optical responses in atomic scale.

This paper is organized as follows.
In section \ref{sec:theory}, we provide a formalism and a computational method based on TDDFT
to extract the individual component of the nonlinear polarization. Calculated nonlinear polarization
components in time domain and their analyses are presented in section \ref{sec:result}.
Finally, in section \ref{sec:summary}, a summary will be presented.

\section{Formalism}
\label{sec:theory}

\subsection{Time-dependent Kohn-Sham equation}

In this section, we explain our formalism to calculate nonlinear polarization in time domain
and to decompose it into perturbative series.
In optical frequencies, the applied electric field can be treated as spatially uniform in
a unit cell of crystalline solids (dipole approximation) since the wavelength is much longer
than both the spatial scale of the electron motion induced by the field and the lattice constant of the cell.
Therefore, we describe the electron dynamics using the following TDKS equation in a unit cell of crystalline solids
\cite{bertsch2000real, otobe2008first, shinohara2010coherent}:
\begin{align}
  i\hbar \frac{\partial}{\partial t}
  u_{ n \mathbf{k}}(\mathbf{r}, t)
  =&
  \left[ 
  \frac{1}{2m} \left(
  \hat{\mathbf{p}} + \hbar\mathbf{k}
  +
  \frac{e}{c}{\mathbf{A}(t)}
  \right)^2
  \right.
  \notag \\
  & \left.
  +
  V_{\mathrm{ion}}
  +
  V_{\mathrm{H}}
  +
  V_{\mathrm{xc}}
  \right]
  u_{n \mathbf{k}}(\mathbf{r}, t)
  \;,
  \label{eq:tdks}
\end{align}
where $e$, $m$, $\hbar$ are the elementary charge, electron mass, and reduced Planck constant,
respectively.
$u_{ n \mathbf{k}}(\mathbf{r},t)$ is the Bloch orbital specified by the crystalline wave number
$\mathbf{k}$ and the band index $n$.
The vector potential $\mathbf{A}(t)$ is related to the applied pulsed electric field by $\mathbf{E}(t) = - (1/c) [\partial\mathbf{A}(t)/\partial{t}]$.
$V_\mathrm{ion}$, $V_{\mathrm{H}}$ and $V_{\mathrm{xc}}$ are the ionic (pseudo-) potential, the Hartree potential and
the exchange-correlation potential, respectively. In the present calculation, we ignore exchange-correlation effects
on the vector potential for simplicity.

From the Bloch orbitals, the electric current density $\mathbf{J}(t)$ is obtained as below:
\begin{align}
  \mathbf{J}(t)
  =
  - \frac{e}{m\Omega}
  \sum_{n \mathbf{k}}
  & \left[
    \int {
      u_{ n \mathbf{k}}^*(\mathbf{r}, t)
      \left(
        \hat{\mathbf{p}} + \hbar\mathbf{k} + \frac{e}{c}{\mathbf{A}(t)}
      \right)
      u_{ n \mathbf{k}}(\mathbf{r}, t)
      \;
      \mathrm{d} \mathbf{r}
    }
  \right.   \notag \\
  & + \frac{m}{i \hbar} 
  \left. \int {
    u_{ n \mathbf{k}}^*(\mathbf{r}, t)
    \left[\hat{\mathbf{r}}, V_{\mathrm{NL}}\right]
    u_{ n \mathbf{k}}(\mathbf{r}, t)
    \;
    \mathrm{d} \mathbf{r}
  }
  \right] 
  \label{eq:j}
  \;,
\end{align}
where $\Omega$ is the volume of the unit cell, $V_\mathrm{NL}$ is the nonlocal part of the pseudo-potential $V_\mathrm{ion}$.
The induced polarization density $\mathbf{P}(t)$ is obtained by integrating the electric
current density over time:
\begin{align}
  \mathbf{P}(t)
  =
  \int^t {
  \mathbf{J}(t')
  \; \mathrm{d}t'
  }
  \label{eq:p}
  \;.
\end{align}

\subsection{Perturbative Expansion of Nonlinear Polarization}
\label{sec:expand_p}

The polarization defined in Eq.~(\ref{eq:p})  contains both linear and nonlinear components.
We numerically decompose it into power series with respect to the field amplitude.
We assume a linearly-polarized pulsed electric field $\mathbf{E}_i(t)$ of the following form,
\begin{align}
  \mathbf{E}_i(t) = E_i \mathbf{e} f(t)
  \;,
  \label{eq:pulse}
\end{align}
where $f(t)$ specifies the time profile of the field that has a maximum value of unity at around $t=0$.
$E_i$ specifies the maximum amplitude of the electric field. $\mathbf{e}$ is a unit vector that specifies
the polarization direction. We will later specify the practical profile of $f(t)$ to be used in the calculations.

To decompose the polarization into power series, we carry out electron dynamics calculations
utilizing the pulsed electric field of the same time profile $f(t)$ and the polarization direction
$\mathbf{e}$, but different maximum amplitudes $E_i$.
We denote the induced polarization caused by the pulsed electric field $\mathbf{E}_i(t)$ as $\mathbf{P}_i(t)$.

Assuming that the applied electric field is sufficiently weak, the induced polarizations $\mathbf{P}_i(t)$ can
be expressed in power series,
\begin{align}
  \mathbf{P}_i(t)
  =
  \sum_n
  \mathbf{p}^{(n)}(t) \left( E_i \right)^n
  \;,
  \label{eq:expand_p}
\end{align}
where $\mathbf{p}^{(n)}(t)$ is the $n$-th order component of the polarization.
When the polarization vector $\mathbf{e}$ coincides with one of Cartesian directions,
$\mathbf{e} = \mathbf{e}_{\alpha}$ $(\alpha=x,y,z)$, the Cartesian components of $\mathbf{p}^{(n)}$ may
be expressed using susceptibility tensors $\chi^{(n)}$ as,
\begin{align}
  p^{(n)}_{\beta}(t)
  =
  \int
  &
  \chi^{(n)}_{\beta \alpha \cdots \alpha}(t-t_1, t-t_2, \cdots t-t_n)
  \notag \\
  &
  \times f(t_1) f(t_2) \cdots f(t_n)
  \;
  \mathrm{d}t_1 \mathrm{d}t_2 \cdots \mathrm{d}t_n
  \;.
  \label{eq:p_chi}
\end{align}
Expressions for a general polarization vector will be obvious.

Once we obtain a set of $N$ results of the polarization amplitudes $\left\{ \mathbf{P}_1(t), \mathbf{P}_2(t), \cdots, \mathbf{P}_N(t)\right\}$
for the field amplitudes $\left\{ E_1, E_2, \cdots, E_N \right\}$, we use Eq.~(\ref{eq:expand_p}) to obtain
$\mathbf{p}^{(n)}(t)$ up to $N$-th order.  Regarding Eq.~(\ref{eq:expand_p}) as a linear system, we have
\begin{align}
  \begin{pmatrix}
    \mathbf{p}^{(1)}(t) \\
    \vdots \\
    \mathbf{p}^{(N)}(t)
  \end{pmatrix}
  =&
  \begin{pmatrix}
    E_1 & \cdots & E_1^{N} \\
    \vdots & \ddots & \vdots \\
    E_N & \cdots & E_N^{N}
  \end{pmatrix}^{-1}
  \begin{pmatrix}
    \mathbf{P}_1(t) \\
    \vdots \\
    \mathbf{P}_N(t)
  \end{pmatrix}
  \;.
  \label{eq:decompose_p}
\end{align}
To carry out the inversion accurately and stably, the following conditions should be satisfied:
\begin{enumerate}
  \item The amplitude $E_i$'s should be sufficiently small so that components higher than $N$-th order
  included in $\mathbf{P}_i(t)$ can be ignored.
  \item The amplitude $E_i$'s should be sufficiently large so that the extracted nonlinear components
  do not suffer substantial numerical noises and/or rounding errors.
\end{enumerate}
We can find an appropriate amplitude window to carry out the stable inversion procedure,
as will be described below.

In the calculations presented in later sections, we will extract nonlinear polarization components
up to third order and only for cases where the inversion symmetry exists and  the second-order component,
$\mathbf{p}^{(2)}(t)$, vanishes identically. For such cases, two calculations with the maximum amplitudes,
$E_1$ and $E_2$, should be sufficient to extract two components, $\mathbf{p}^{(1)}(t)$ and $\mathbf{p}^{(3)}(t)$,
In practice, however, we find it useful to calculate for the phase-inverted pair, the time evolution calculation
using the electric field with the sign inverted, to remove the even-order components accurately.
We carry out four calculations using electric fields of ($\mathbf{E}_1(t) $ and $\mathbf{E}_2(t)$) with the
maximal amplitudes of $E_1, E_2$, and ($\overline{\mathbf{E}_1}(t)$ and $\overline{\mathbf{E}_2}(t)$)
that are their phase-inverted pairs.
The induced polarizations corresponding to these four applied pulses are denoted as
$\mathbf{P}_1(t)$, $\mathbf{P}_2(t)$, $\overline{\mathbf{P}}_1(t)$ and  $\overline{\mathbf{P}}_2(t)$, respectively.
Using them, the solution of the linear system of Eq. (\ref{eq:decompose_p}) is expressed as
\begin{align}
  \mathbf{p}^{(1)}(t)
  =&
  \frac{
  E_1^3 \left[
  {\mathbf{P}}_2(t)
  -
  \overline{\mathbf{P}}_2(t)
  \right]
  -
  E_2^3 \left[
  {\mathbf{P}}_1(t)
  -
  \overline{\mathbf{P}}_1(t)
  \right]
  }{
  2 E_1 E_2 (E_1 - E_2) (E_1 + E_2)
  }
  \label{eq:p1}
  \\
  \mathbf{p}^{(3)}(t)
  =&
  -
  \frac{
  E_1 \left[
  {\mathbf{P}}_2(t)
  -
  \overline{\mathbf{P}}_2(t)
  \right]
  -
  E_2 \left[
  {\mathbf{P}}_1(t)
  -
  \overline{\mathbf{P}}_1(t)
  \right]
  }{
  2 E_1 E_2 (E_1 - E_2) (E_1 + E_2)
  }
  \label{eq:p3}
  \;.
\end{align}
This formula will be used in the calculations shown below.

\subsection{Expansion of Nonlinear Charge Density}
\label{sec:expand_rho}

The power-series inversion method can also be applied to other observables.
We will apply the method to electron density $\rho(\mathbf{r}, t)$ as follows:
\begin{align}
  \rho(\mathbf{r}, t) - \rho_\mathrm{GS}(\mathbf{r})
  =&
  \sum_{n}
  \rho^{(n)}(\mathbf{r}, t) \left( E_i \right)^n,
\end{align}
where $\rho(\mathbf{r},t)$ and  $\rho_\mathrm{GS}(\mathbf{r})$ represent the electron density at time $t$
and that in the ground state, respectively.
Using a similar inversion procedure to Eq.~(\ref{eq:decompose_p}), perturbative components of
electron density change, $\rho^{(n)}(\mathbf{r},t)$, can be determined.
We will show that such decomposition is useful to understand the nonlinear interaction in microscopic
scale and to investigate the physical origin of the nonlinear polarization.

\section{Results and Discussion}
\label{sec:result}

\begin{figure}
  \includegraphics[width=0.9\linewidth]{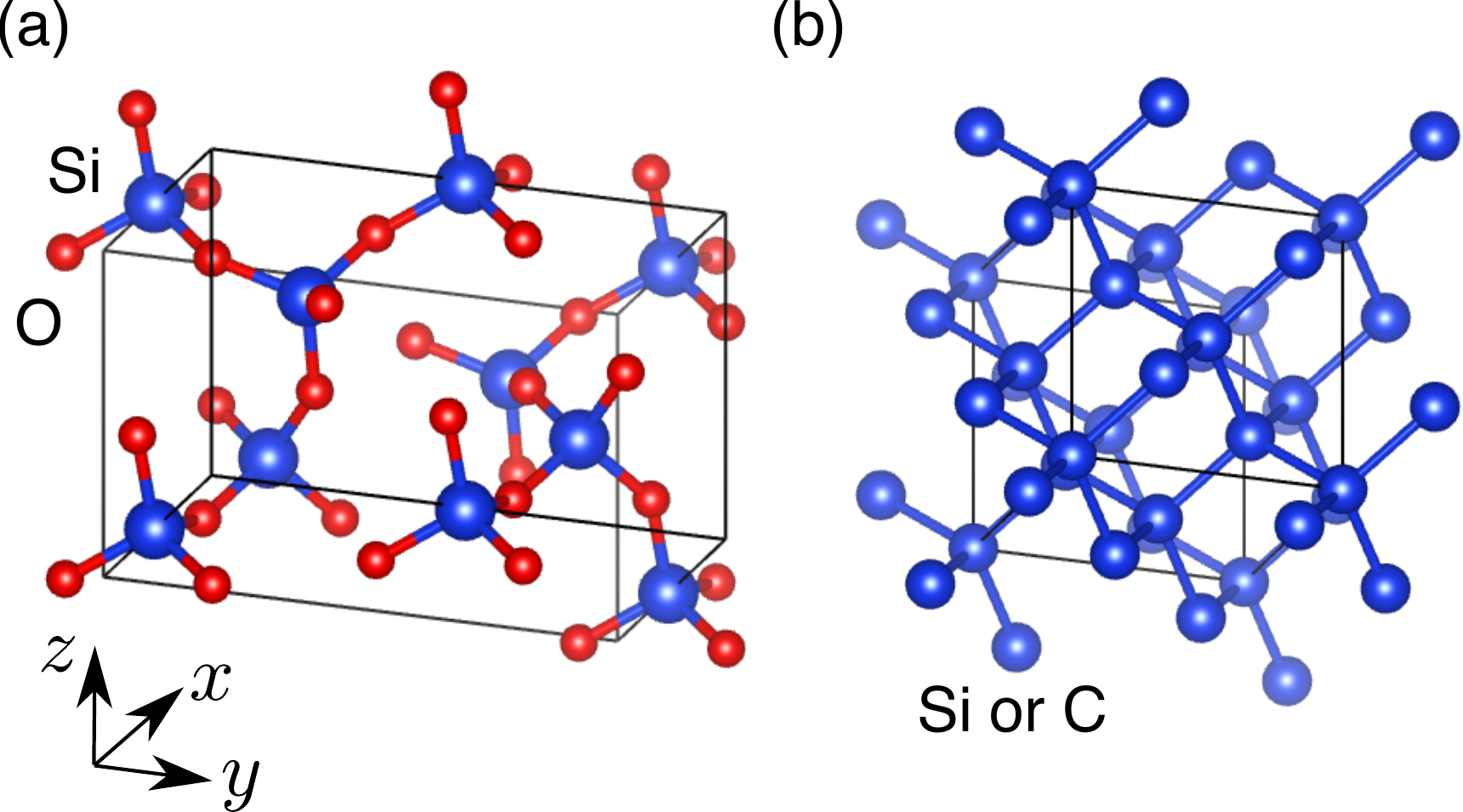}
  \caption{Crystal structure of $\alpha$-$\mathrm{SiO}_2$ (a) and silicon/diamond (b).}
  \label{fig:structure}
\end{figure}

We apply our method to three different bulk materials;  $\alpha$-$\mathrm{SiO}_2$, diamond ($\mathrm{C}$),
and silicon ($\mathrm{Si}$). We show their crystal structures and the unit cells used in our calculations in
Fig.~\ref{fig:structure}(a) and (b).
$\alpha$-$\mathrm{SiO}_2$ is an ionic crystal with a wide optical gap.
Diamond and silicon are typical covalent crystals with the diamond structure, having different optical gap energies.

For the time profile of the applied electric field, we adopt the one defined for the vector potential:
\begin{equation}
  \mathbf{A}(t)
  =
  {A}_i
  \mathbf{e}
  \cos^2 \left\{ \frac{\pi}{T_L} \left( t - \frac{T_L}{2} \right) \right\}
  \sin \omega_L \left( t-\frac{T_L}{2} \right),
\end{equation}
for $0 < t < T_L$ and 0 for otherwise.
$T_L$ and $\omega_L$ are the pulse length and the central frequency of the applied electric field, respectively.
The maximal amplitude of the vector potential $A_i$ is related to the maximal electric field amplitude $E_i$ by $E_i=(\omega_L/c)A_i$.
Numerical values used in our calculations are summarized in Table~\ref{tab:pulse}.

\begin{table}
  \caption{Parameters of applied electric fields}
  \label{tab:pulse}       
  \begin{tabular}{lll}
    \hline\noalign{\smallskip}
    Central Frequency $[\mathrm{eV}]$ & $\hbar\omega_L $ & 1.55  \\
    \noalign{\smallskip}\noalign{\smallskip}
    Pulse Length $[\mathrm{fs}]$ & $T_p$ & 20  \\
    \noalign{\smallskip}\noalign{\smallskip}
    Amplitude [{V}/\AA]
    & $E_1$ & $8.680\times 10^{-2}$  \\
    & $E_2$ & $6.138\times 10^{-2}$  \\
    & $E_3$ & $3.882\times 10^{-2}$  \\
    & $E_4$ & $2.745\times 10^{-2}$  \\
    & $E_5$ & $8.680\times 10^{-3}$  \\
    \noalign{\smallskip}\hline
  \end{tabular}
\end{table}

As for the exchange-correlation potential $V_{\mathrm{xc}}$ in Eq.~(\ref{eq:tdks}),
we assume an adiabatic approximation utilizing the same functional form of the potential
as that used in the ground state calculation.
We will employ two different types for the potential. One is the local density approximation (LDA) \cite{perdew1981self}.
As is well known, optical gap energies are substantially underestimated in the LDA. Related to this failure,
$\chi^{(1)}$ values are also often overestimated.
The other one is TBmBJ potential \cite{tran2009accurate}. It depends on kinetic energy density and belongs to
a functional group of meta generalized gradient approximation.
The TBmBJ potential is known to give reasonable descriptions for the bandgap energies of various dielectrics.
We choose the parameter of the TBmBJ potential, $c_m$, to reproduce the experimental optical gap energies.

For numerical calculations, we use an open-source TDDFT program package, SALMON
(Scalable Ab-initio Light-Matter simulator for Optics and Nanoscience),
which has been developed in our group \cite{noda2018salmon}.
Before starting the time evolution calculation, we first calculate the ground state which will be used as an initial state
of the time evolution calculations.
We then carry out time evolution calculations of Bloch orbitals solving Eq.~(\ref{eq:tdks}) in time domain.
The current density and the polarization are calculated according to Eqs. (\ref{eq:j}) and (\ref{eq:p}).
To express Bloch orbitals,  we use a uniform grid system in the three-dimensional Cartesian coordinates.
The Brillouin zone is uniformly sampled by the Monkhorst-Pack grids.
The numerical parameters used in the calculations are summarized in Table \ref{tab:param}.

\begin{table}
  \caption{Numerical parameters used in the calculations.
  Atomic unit is used for length and time.}
  \label{tab:param}       
  \begin{tabular}{llll}
    \hline\noalign{\smallskip}
    & Lattice constant & Time step & \\
    & $r$-space grid &  $k$-space grid   \\
    \noalign{\smallskip}\hline\noalign{\smallskip}
    $\mathrm{Si}$ & $10.26 \times 10.26 \times 10.26$ & $0.08$ & \\
    & $16 \times 16 \times 16$ & $16 \times 16 \times 16$   \\
    \noalign{\smallskip}\noalign{\smallskip}
    $\alpha$-$\mathrm{SiO}_2$ & $9.28 \times 16.1 \times 10.2$ & $0.04$ & \\
    & $20 \times 36 \times 52$ & $4 \times 4 \times 4$   \\
    \noalign{\smallskip}\noalign{\smallskip}
    C & $6.74 \times 6.74 \times 6.74$ &  $0.02$ & \\
    & $24 \times 24 \times 24$ & $12 \times 12 \times 12$   \\
    \noalign{\smallskip}\hline
  \end{tabular}
\end{table}

\subsection{$\alpha$-$\mathrm{SiO}_2$}

We  extract nonlinear polarization components $\mathbf{p}^{(n)}(t)$ for $\alpha$-SiO$_2$
using the procedure explained in Sec.~\ref{sec:expand_p}.
We choose the polarization direction parallel to $c$-axis which is set to coincide with the $z$-direction.
In this setting, the direction of the polarization is parallel to the direction of the applied electric field.
There appear no second-order polarization due to the crystal symmetry.

\begin{figure}
  \includegraphics[width=0.9\linewidth]{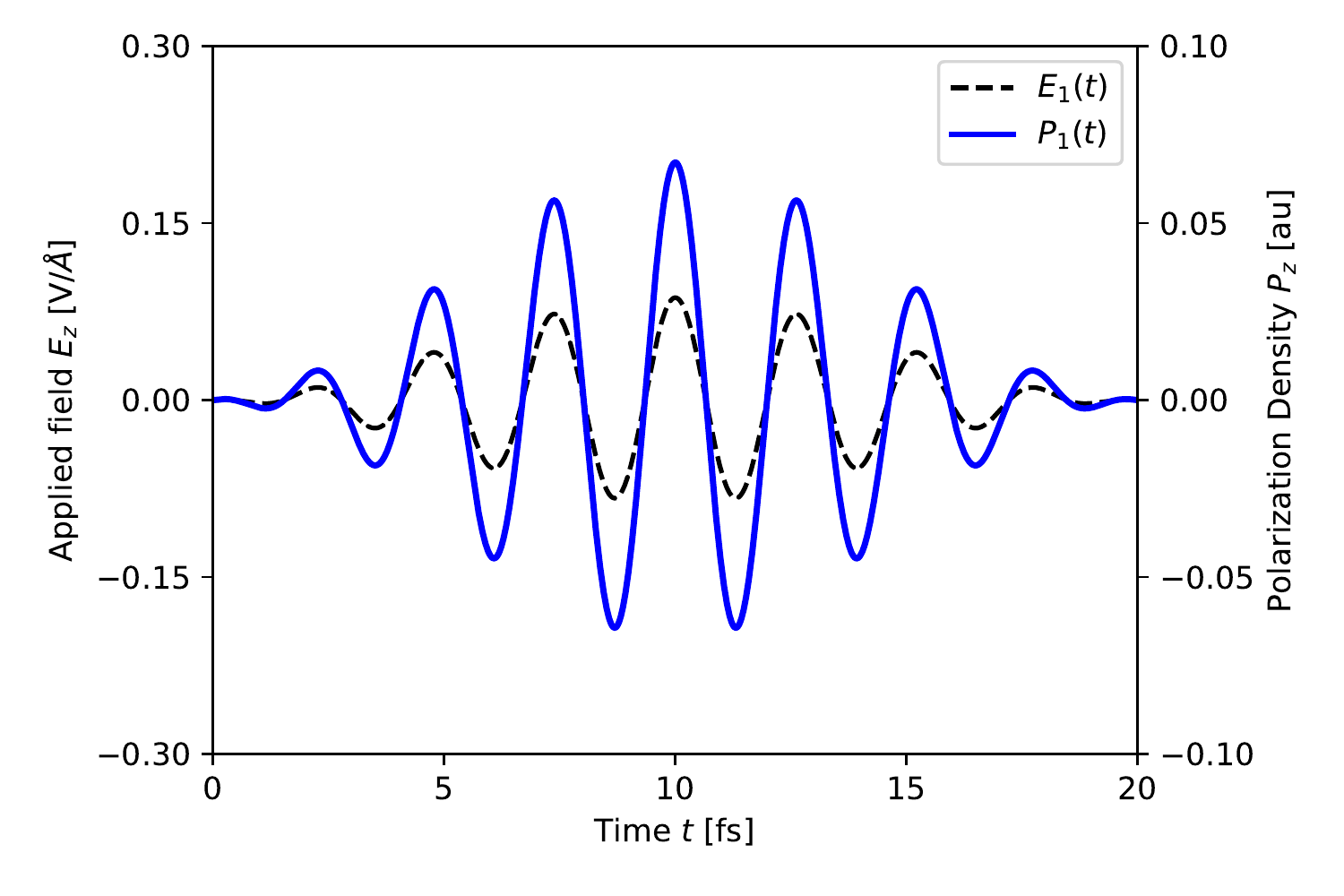}
  \caption{
  Applied electric field (red-dashed line) and the induced polarization (blue-solid line) in $\alpha$-$\mathrm{SiO}_2$.
  }
  \label{fig:sio_ep}
\end{figure}

First we show the results employing the adiabatic LDA.
Figure~\ref{fig:sio_ep} show the applied electric field and the induced polarization.
The maximam amplitude of the field is set to $E_1 = 8.680\times 10^{-2}$ $\mathrm{V}/\mathrm{\AA}$,
the weakest one among those listed in Table 2.
$P(t)$ looks proportional to $E(t)$, since the response is dominated by the linear polarization.

\begin{figure}
  \includegraphics[width=0.9\linewidth]{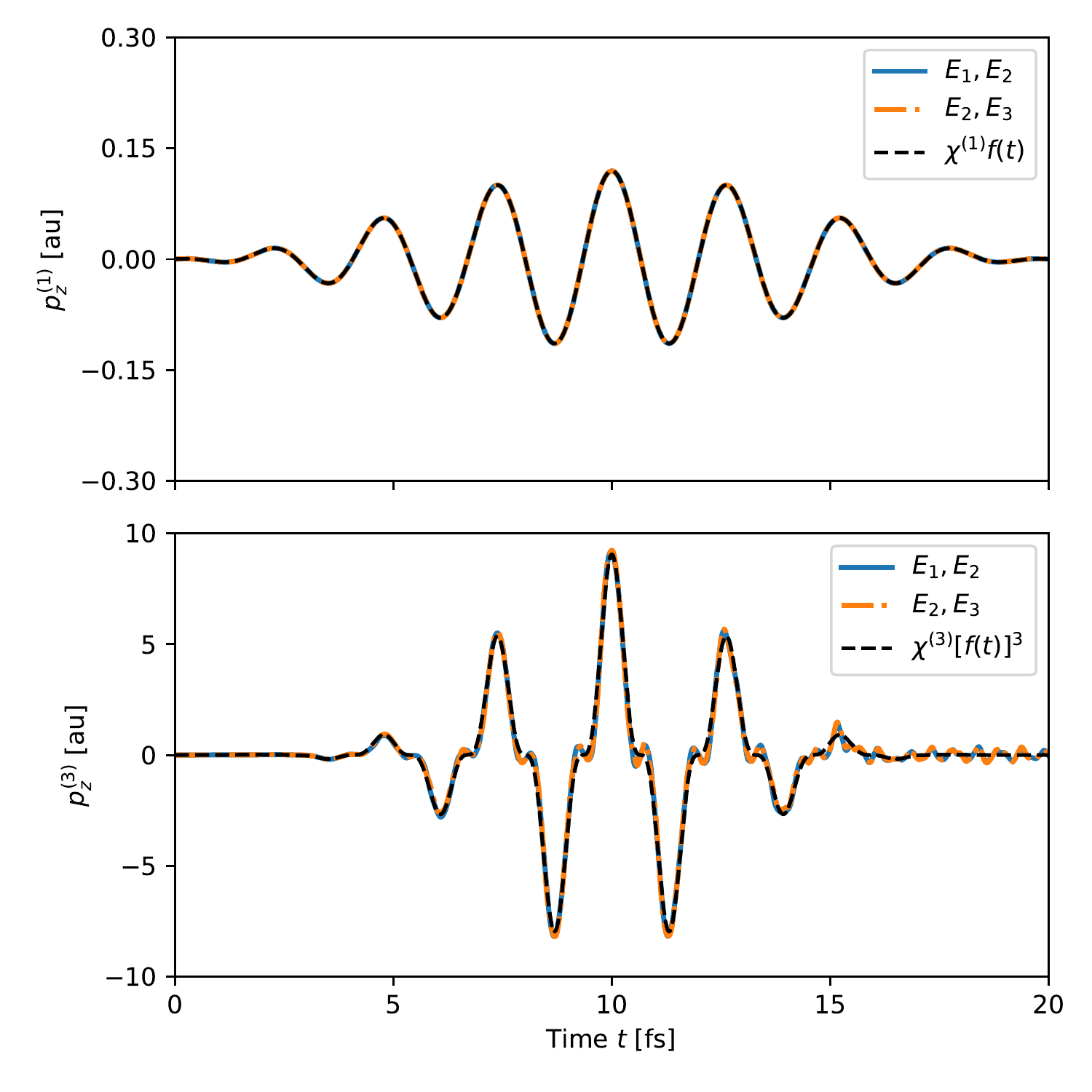}
  \caption{(a) Linear component, $p^{(1)}$, and (b) third-order nonlinear component, $p^{(3)}$, of the
  induced polarization in $\alpha$-SiO$_2$. The blue-solid and the orange-dash-dot lines are calculated using a set of $E_1, E_2$ and
  a set of $E_2, E_3$, respectively. Black-dashed lines are the fit using simple functions. See text for the detail.
  }
  \label{fig:sio_epn}
\end{figure}

In Fig.~\ref{fig:sio_epn}, we show calculated ${p}^{(1)}(t)$ and ${p}^{(3)}(t)$.
Results using two amplitudes of $E_1$ and $E_2$ are shown by blue-solid lines, and
those using $E_2$ and $E_3$ are shown by orange-dash-dot lines.
An excellent agreement of the two calculations indicates that our procedure to decompose
the polarization into power series works with high accuracy.

As a next step, we fit the polarization components using simple functional forms,
$f(t)$ for linear and $f(t)^3$ for third-order nonlinear components.
\begin{align}
  p^{(1)}(t) =& \chi^{(1)} f(t) \label{eq:p1_chi1}\\
  p^{(3)}(t) =& \chi^{(3)} \left[f(t)\right]^3 \label{eq:p3_chi3}
  \;.
\end{align}
As shown by black-dashed lines in Fig.~\ref{fig:sio_epn}, the fitting works very well.
Since the optical gap of $\alpha$-$\mathrm{SiO}_2$ is much larger than the three-photon
energy ($3\hbar\omega_1 = 4.65$~eV), real electron-hole excitations do not take place up
to the third-order nonlinearities. This explains the reason why we can fit the polarization
components ignoring the retardation effects.
In panel (b), a small amplitude, high frequency oscillation is seen after the pulse ends.
We consider this oscillation comes from real electronic excitations caused by small
high-frequency components that are included in the applied electric field.

\begin{table}[tb]
  \begin{tabular}{lllll}
    \hline
    TDDFT
    & LDA  & $\chi^{(1)}$ & $1.50$ \\
    &     & $\chi^{(3)}$ & $4.3\times10^{-22}$~($\mathrm{m}^2/\mathrm{V}^2$) \\
    &     & $E_G$ & $6.3$~eV \\
    \noalign{\smallskip}\noalign{\smallskip}
    & TBmBJ & $\chi^{(1)}$ & $1.18$ \\
    &     & $\chi^{(3)}$ & $1.3\times10^{-22}$~($\mathrm{m}^2/\mathrm{V}^2$) \\
    &     & $E_G$ & $8.9$~eV ($c_m=1.2$) \\
    \noalign{\smallskip}\noalign{\smallskip}
    Exp. \cite{adair1989nonlinear}
    & & $\chi^{(1)}$ & $1.38$ \\
    & & $\chi^{(3)}$ & $2.65\times10^{-22}$~($\mathrm{m}^2/\mathrm{V}^2$)  \\

    \hline
  \end{tabular}
  \caption{
  Calculated values of $\chi^{(1)}$ and $\chi^{(3)}$ for $\alpha$-$\mathrm{SiO}_2$.
  }
  \label{tab:sio2}
\end{table}

In Table~\ref{tab:sio2}, extracted nonlinear susceptibilities employing LDA and TBmBJ
functionals are shown and are compared with measured values.
Calculations using LDA overestimate both $\chi^{(1)}$ and $\chi^{(3)}$ values.
Employing TBmBJ, $\chi^{(1)}$ decreases by a factor of 3. For both quantities, the measured values lie between two calculations.

\begin{figure*}
  \resizebox{0.9\textwidth}{!}{
  \includegraphics[width=1.0\linewidth]{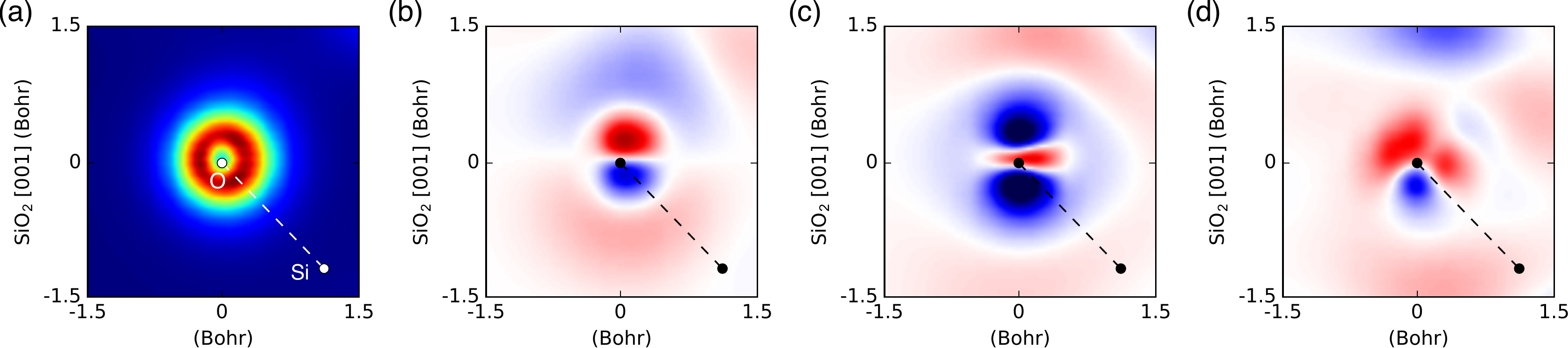}
  }
  \caption{
  Electron density profiles on a slice containing both $z$-axis and a $\mathrm{Si-O}$ bond
  are decomposed into power series with respect to the field amplitude \cite{sato2016thesis}.
  (a) Ground state electron density $\rho_\mathrm{GS}$.
  (b-d) 1st, 2nd and 3rd-order components of the electron density.
  }
  \label{fig:sio_rho}       
\end{figure*}

We next show a decomposition of electron density distribution in powers of the electric field amplitude using the method explained in Sec.~\ref{sec:expand_rho}.
We expect such decomposition provides a useful information regarding the microscopic origin of the nonlinear polarization components \cite{sato2016thesis}. 
Figure~\ref{fig:sio_rho} shows the results.
As seen from (a), valence electrons locate dominantly around O atoms, while electron density around Si atoms is small in the ground state.
The linear, the second-order nonlinear, and the third-order nonlinear components, $\rho^{(1)}$, $\rho^{(2)}$ and $\rho^{(3)}$, are
shown in (b), (c), and (d), respectively, at a time when the electric field is maximum.
The red and the blue colors indicate the positive (increase from the ground state) and the negative (decrease from the ground state)
changes in the density.

The linear component shown in (b) is dominated by the dipole motion of electrons around O atoms.
The second-order nonlinear component shown in (c) is dominated by the quadrupole motion around O atoms.
Finally, the third-order nonlinear component in (d) shows a complex behavior. While dipole-like shape is seen around
the O atoms, a contribution from electrons in more distant place than the case of the linear response is found.
In any case, both linear and nonlinear optical responses are dominantly originated from the electron dynamics around
the O atoms, since electrons locate dominantly around them in the ground state.

\subsection{Diamond ($\mathrm{C}$)}

\begin{figure}
  \includegraphics[width=0.9\linewidth]{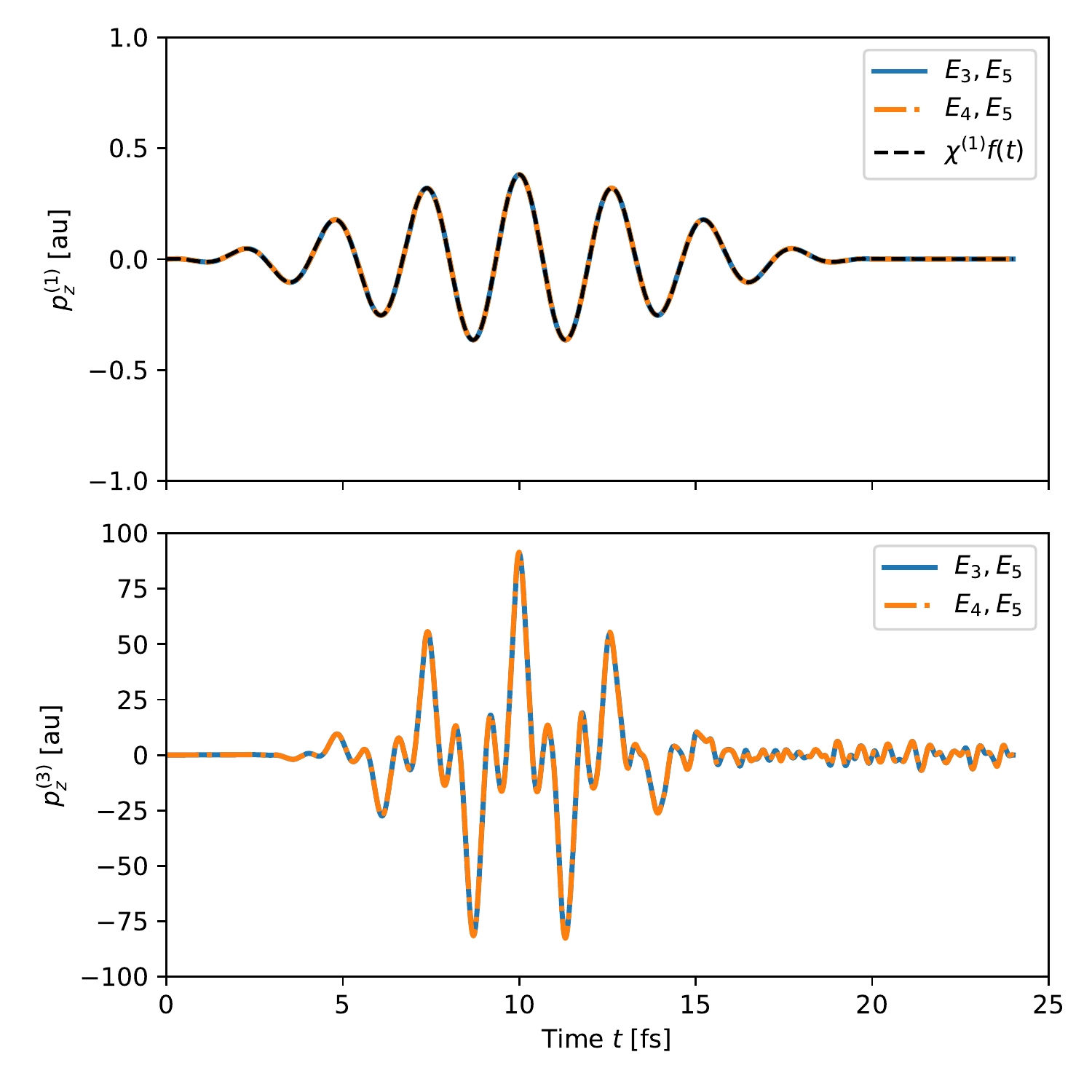}
  \caption{
  Polarization components of diamond extracted by our calculations.
  (a) Linear component, $p^{(1)}(t)$.
  (b) Nonlinear (3rd-order) component, $p^{(3)}(t)$.
  The blue-solid and the orange-dashed curves are calculated using the sets of the electric field amplitude of $E_3, E_5$ and $E_4, E_5$, respectively.
  }
  \label{fig:c_pn}
\end{figure}

Diamond ($\mathrm{C}$) is a covalent material with a large band gap energy.
We extract linear and nonlinear polarization components using the same
procedure as that used for $\alpha$-SiO$_2$.
We apply the pulsed electric field along one cubic axis.
The result employing adiabatic LDA is shown in Fig. \ref{fig:c_pn}.
Calculations using two sets of different amplitudes, ($E_3$, $E_5$) and
($E_4$, $E_5$), again coincide accurately with each other.
This confirms that the extraction of the third-order nonlinear component
is accurately carried out.
From the time profile of the third-order nonlinear component shown in Fig. \ref{fig:c_pn}(b),
it is apparent that the time profile is not simply proportional to $f(t)^3$,
contrarily to the case of $\alpha$-SiO$_2$.

Since the applied electric field dominantly contains frequency components around $\omega_L$,
the third-order component $p^{(3)}(t)$ should dominantly contain two frequency components
of $\omega_L$ and $3\omega_L$. They are related to the optical Kerr effect and the
third harmonic generation (THG), respectively.
To examine each frequency component separately, we perform a frequency-domain
filtering of $p^{(3)}(t)$, decomposing it into low ($|\omega| < 2 \omega_L$)
frequency component, $p^{(3)}(t; \omega_L)$, and high ($2\omega_L
< |\omega| < 4\omega_L$) frequency component, $p^{(3)}(t;3\omega_L)$,
as defined by
\begin{align}
  p^{(3)}(t; \omega_L)
  =&
  \int_{-2\omega_L}^{2\omega_L}
  \mathrm{d}\omega
  p^{(3)}(\omega)
  e^{-i\omega t},
  \label{eq:p3z_low}
  \\
  p^{(3)}(t; 3\omega_L)
  =&
  \left(
  \int_{-4\omega_L}^{-2\omega_L}
  +
  \int_{2\omega_L}^{4\omega_L}
  \right)
  \mathrm{d}\omega\;
  p^{(3)}(\omega)
  e^{-i\omega t},
  \label{eq:p3z_high}
\end{align}
with
\begin{align}
  p^{(3)}(\omega)
  =&
  \frac{1}{2\pi}
  \int_{0}^{T}
  \mathrm{d}t'\;
  e^{i\omega t'} p^{(3)}(t').
  \label{eq:p3z_ft}
  \;
\end{align}

\begin{figure}
  \includegraphics[width=0.9\linewidth]{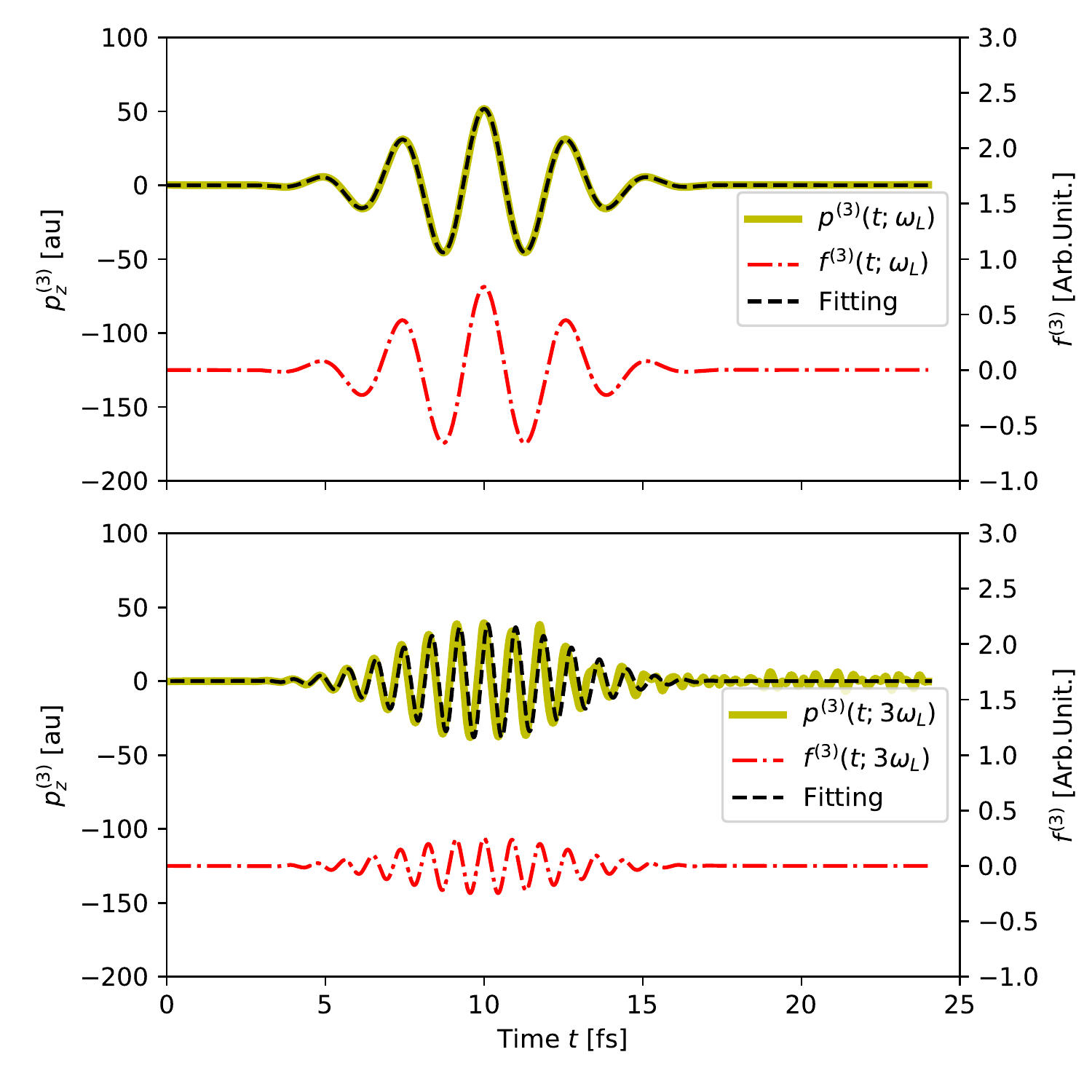}
  \caption{
  Fourier-filtered 3rd-order components of the nonlinear polarization (yellow curves) and those of the time profile of the applied pulse cubed, $f(t)^3$,
  (red-dashed curves), (a) for low- and (b) for high-frequencies. Black-dashed curves show fitting results using Eqs.~(\ref{eq:p3_w1d0}) and (\ref{eq:p3_w3d0}).
  }
  \label{fig:c_p3_spec}
\end{figure}

In Fig. \ref{fig:c_p3_spec}, we show by yellow-solid curves the low frequency component $p^{(3)}(t; \omega_L)$
in (a) and the high frequency component $p^{(3)}(t;3\omega_L)$ in (b).
We also show filtered functions of $f(t)^3$ by red-dash-dotted curves:  the low frequency component
$f^{(3)}(t;\omega_L)$ in (a) and the high frequency component $f^{(3)}(t;\omega_L)$ in (b).
As seen from the figure, the time profile of the frequency components $p^{(3)}(t;\omega_L)$ looks very similar
$f^{(3)}(t,\omega_L)$, and also for $p^{(3)}(t;3\omega_L)$ and $f^{(3)}(t;3\omega_L)$.

In $p^{(3)}(t;3\omega_L)$, an oscillation with a small amplitude continues after
the applied electric field ends. We consider that this oscillation comes from a
real, resonant excitation by three-photon absorption. In our calculation, the direct
band gap energy of the diamond employing LDA is 4.7 eV, which is slightly larger than
$3\hbar \omega_1 = 4.65$ eV. However, because of the wide spectral distribution
of the applied pulse, excitation across the band gap is partly possible by the three-photon
absorption.

The above analysis shows that, while the time profile of the third-order polarization
$p^{(3)}(t)$ is rather different from the time profile of $f(t)^3$, the filtered functions
have very similar shapes for high and low frequency components.
This fact indicates a strong frequency dependence of the third-order nonlinear
susceptibilities, $\chi^{(3)}(\omega_L) = \chi^{(3)}(\omega_L; \omega_L,\omega_L,-\omega_L)$
and $\chi^{(3)}(3\omega_L) = \chi^{(3)}(3\omega_L; \omega_L, \omega_L, \omega_L)$.
We extract them by simply fitting
\begin{equation}
  p^{(3)}(t;\omega_L) = \chi^{(3)}(\omega_L) f^{(3)}(t;\omega_L),
  \label{eq:p3_w1d0}
\end{equation}
\begin{equation}
  p^{(3)}(t;3\omega_L) = \chi^{(3)}(3\omega_L) f^{(3)}(t;3\omega_L),
  \label{eq:p3_w3d0}
\end{equation}
We assume that there is no time delay in the response.
The result of the fitting is shown by black-dashed curves in Fig.~\ref{fig:c_p3_spec}.

Linear and nonlinear susceptibilities obtained by the analysis using LDA and TBmBJ are
summarized in Table.~\ref{tab:c}.
We find the linear susceptibility $\chi^{(1)}$ is reasonably reproduced by the calculation.
The result employing TBmBJ is in better agreement with the measured value.
As for $\chi^{(3)}(\omega_L)$ and $\chi^{(3)}(3\omega_L)$, the latter is larger than
the former in both LDA and TBmBJ calculations. The values employing TBmBA is about a factor of
two to three smaller than those employing LDA, consistent with the larger gap energy using TBmBJ.
The measured values of $\chi^{(3)}$ are in reasonable agreement with the calculated values using TBmBJ.
Semi ab-initio calculation using tight-binding model \cite{moss1991semi} gives slightly smaller value.

\begin{table}[tb]
  \begin{tabular}{lllll}
    \hline
    TDDFT
    & LDA  & $\chi^{(1)}$ & $4.79$ \\
    &     & $\chi^{(3)}(\omega_L)$ & $3.3\times10^{-21}$~($\mathrm{m}^2/\mathrm{V}^2$) \\
    &     & $\chi^{(3)}(3\omega_L)$ & $7.5\times10^{-21}$~($\mathrm{m}^2/\mathrm{V}^2$) \\
    &     & $E_G$ & $4.7$~eV \\
    \noalign{\smallskip}\noalign{\smallskip}
    & TBmBJ & $\chi^{(1)}$ & $4.87$ \\
    &     & $\chi^{(3)}(\omega_L)$ & $1.6\times10^{-21}$~($\mathrm{m}^2/\mathrm{V}^2$) \\
    &     & $\chi^{(3)}(3\omega_L)$ & $2.7\times10^{-21}$~($\mathrm{m}^2/\mathrm{V}^2$) \\
    &     & $E_G$ & $5.9$~eV ($c_m=1.27$) \\
    \noalign{\smallskip}\noalign{\smallskip}
    Exp.
    & & $\chi^{(1)}$ & $4.86$ \cite{boyd2003nonlinear} \\
    & & $\chi^{(3)}$ & $2.1\times10^{-21}$~($\mathrm{m}^2/\mathrm{V}^2$) \cite{boyd2003nonlinear} \\
    & & & $1.8\times10^{-21}$~($\mathrm{m}^2/\mathrm{V}^2$) \cite{almeida2017nonlinear} \\
    Other theory
    & & $\chi^{(3)}$ & $1.1\times10^{-21}$~($\mathrm{m}^2/\mathrm{V}^2$) \cite{moss1991semi} \\
    \hline
  \end{tabular}
  \label{tab:c}
  \caption{
  Calculated linear and nonlinear susceptibilities of diamond.
  }
\end{table}

\subsection{Silicon ($\mathrm{Si}$)}

Silicon ($\mathrm{Si}$) has the same crystalline structure as diamond with a smaller
optical gap energy.
For this system, we investigate the directional dependence of the nonlinear polarization
for which abundant measurements are available.
From the crystalline symmetry, it is sufficient to calculate nonlinear polarization components
for two different directions to investigate the third-order nonlinear polarization.
We adopt $[001]$ and $[011]$ for it.

From the calculations for the two directions, we extract the linear and nonlinear
polarization components. The linear polarization does not depend on the
direction and is expressed as $p^{(1)}(t)$. The third-order nonlinear polarization component
along the two directions are denoted as $p^{(3)}_{[001]}(t)$ and $p^{(3)}_{[011]}(t)$.
From these two components, we define two third-order nonlinear polarizations:
\begin{equation}
  p^{(3)}_{1111}(t) = p^{(3)}_{[001]}(t),
\end{equation}
and
\begin{equation}
  p^{(3)}_{1122}(t) = \frac{2}{3} \left( p^{(3)}_{[011]}(t) - \frac{1}{2} p^{(3)}_{[001]}(t) \right),
\end{equation}
which are related to the third-order nonlinear susceptibilities of $\chi^{(3)}_{1111}$
and $\chi^{(3)}_{1122}$, respectively.

\begin{figure}
  \includegraphics[width=0.9\linewidth]{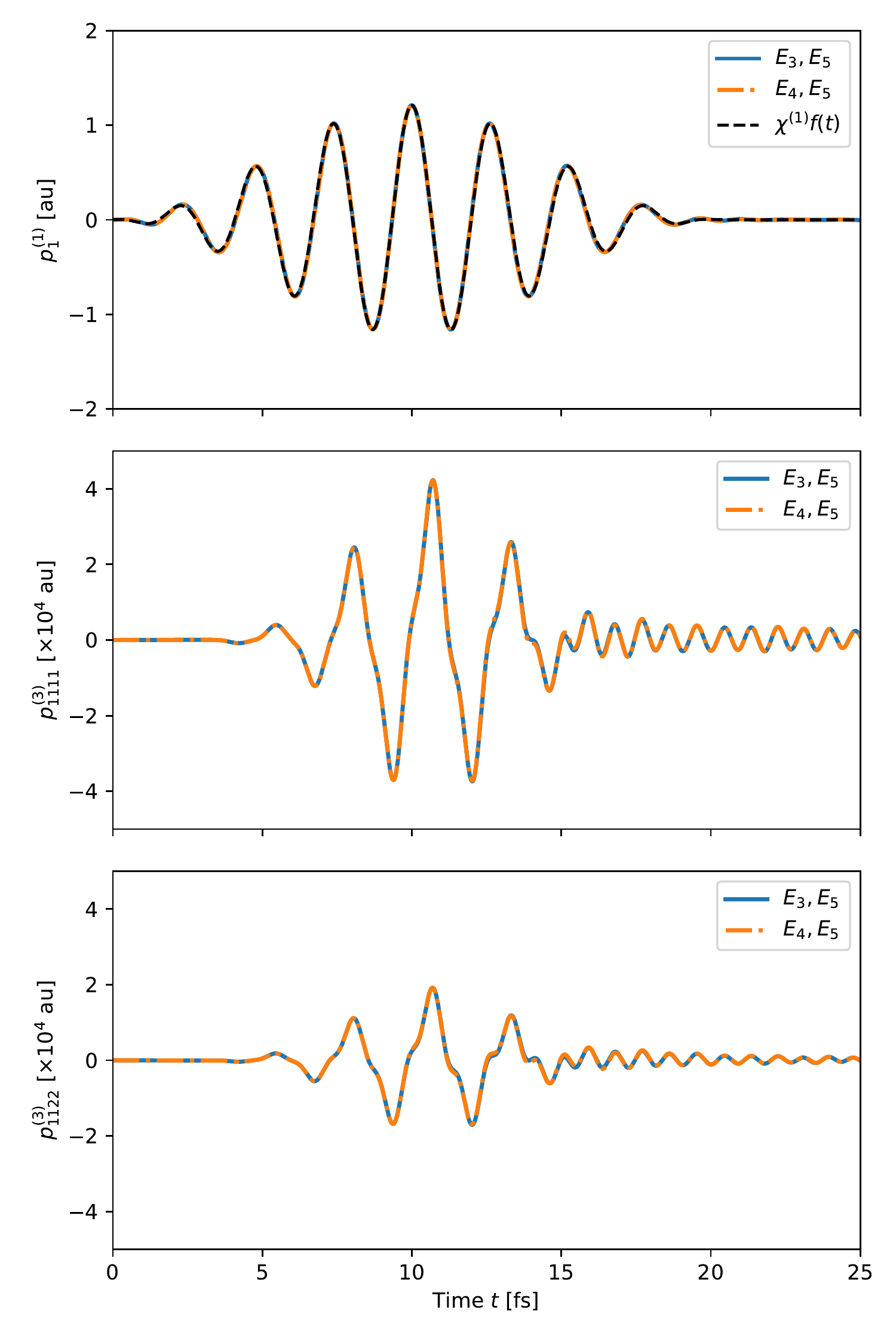}
  \caption{
  Polarization components of silicon extracted by our calculations:
  (a) Linear component, $p^{(1)}(t)$,
  (b) Third-order nonlinear component, $p_{1111}^{(3)}(t)$, and
  (c) $p_{1122}^{(3)}(t)$.
  The blue-solid and the orange-dashed curves are calculated using the sets of the electric field amplitude of $E_3, E_5$ and $E_4, E_5$, respectively.
  }
  \label{fig:si_pn}
\end{figure}

\begin{figure}
  \includegraphics[width=0.9\linewidth]{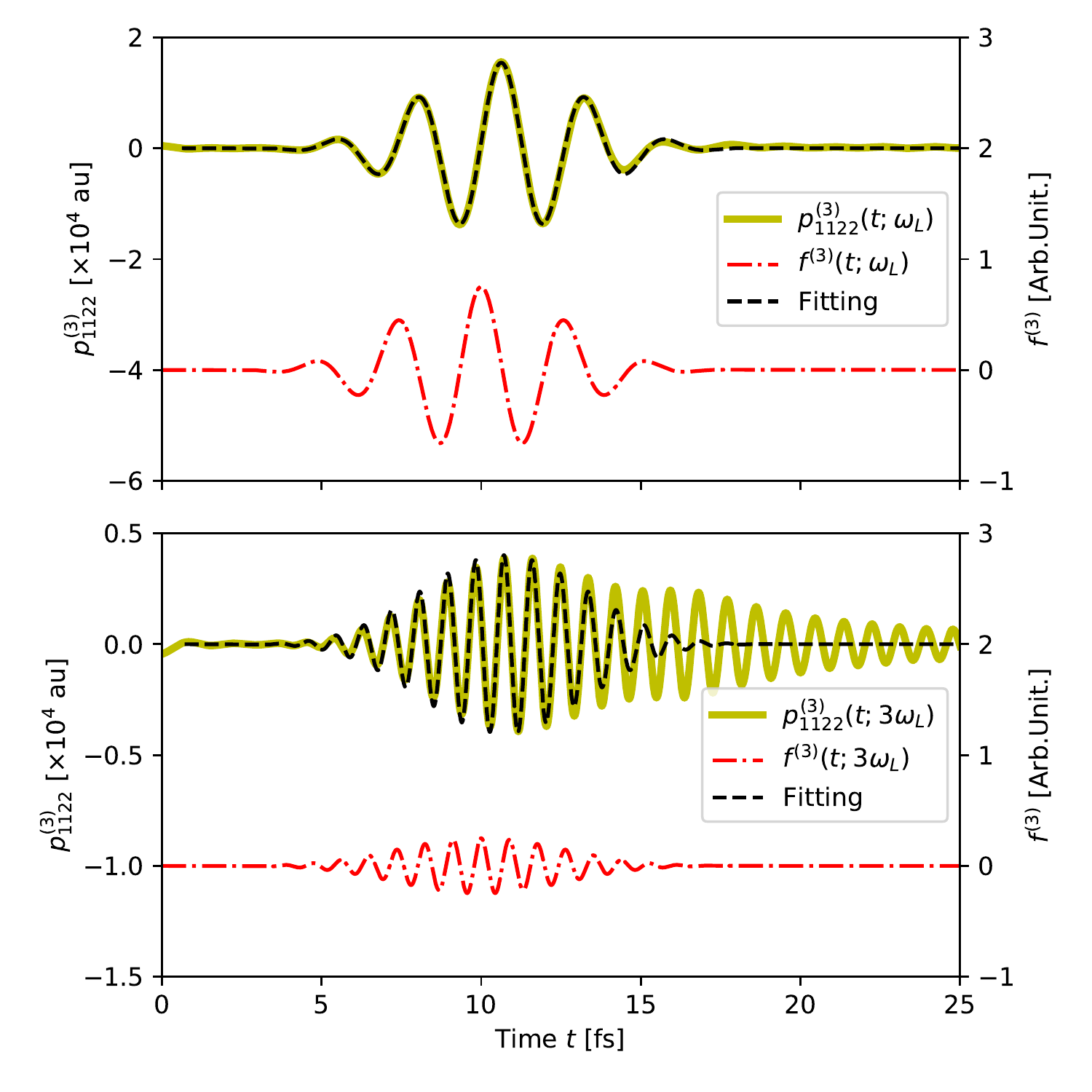}
  \caption{
  Fourier-filtered 3rd-order components of the nonlinear polarization $p^{(3)}_{1122}(t)$ (yellow curves) and those of the time profile of the applied pulse cubed, $f(t)^3$,
  (red-dashed curves), (a) for low- and (b) for high-frequencies. Black-dashed curves are fitting curves using Eq.~(\ref{eq:p3_delay1})-(\ref{eq:p3_delay2}).
  }
  \label{fig:si_p3_spec}
\end{figure}

Fig.~\ref{fig:si_pn} shows $p^{(1)}(t)$, $p^{(3)}_{1111}(t)$, and $p^{(3)}_{1122}(t)$ using pulses
with the amplitudes of $E_3,E_5$ (blue-solid curve)
and $E_4,E_5$ (orange-dotted curve).
Adiabatic LDA is used in the calculation.
A good agreement of two calculations using different sets of amplitudes indicates, again, that the
decomposition of the polarization into nonlinear components works reliably.
Because of the small gap energy, we find it necessary to use electric fields with the amplitude smaller
than those used in diamond and $\alpha$-SiO$_2$ to extract the nonlinear polarization components accurately.

The linear component $p^{(1)}(t)$ can be well fitted using a real constant $\chi^{(1)}$ as $\chi^{(1)}f(t)$ with $\chi_1=1.21$ ($15.2$ in SI units).
The third order components are not simply proportional to the cube of the field, as in the case of diamond.
Employing LDA, the optical gap energy in silicon is about $E_G \sim 2.4$~eV, which is lower than $3\hbar\omega_L \sim 4.65$ eV.
This indicates that real excitations of the valence electrons take place in the third order response.
As seen in Fig.~\ref{fig:si_pn}(b), an oscillating current persists even after the pulse ends.
This also indicates the formation of the excited carriers.
We should note, however, that such oscillatory current is expected to be damped by mechanisms that are not
taken into account sufficiently in the present calculation, for example, dephasing effect caused by the coupling with phonons.

We carry out frequency filtering for the third-order nonlinear components using Eqs. (\ref{eq:p3z_low}), (\ref{eq:p3z_high}), and (\ref{eq:p3z_ft}),
and show results for $p^{(3)}_{1122}(t)$ in Fig.~\ref{fig:si_p3_spec}.
We show the low frequency component $p^{(3)}_{1122}(t;\omega_L)$ in (a) and
the high frequency component $p^{(3)}_{1122}(t;3\omega_L)$ in (b) by yellow-solid curves.
We also show functions applying the filtering procedure to $f(t)^3$, the low frequency component
$f^{(3)}(t;\omega_L)$ in (a) and the high frequency component $f^{(3)}(t;3\omega_L)$ in (b) by red-dash-dot curves.
It is observed that there are phase difference in both $\omega_L$ and $3\omega_L$ components.
The phase difference between $p^{(3)}(t;\omega_L)$ and $f^{(3)}(t,\omega_L)$ is expected, since
the phase difference is necessary for the energy transfer from the applied electric field to electrons in the medium.

We fit the filtered components using functions incorporating the delay time $\delta{t}(\omega_L)$ and $\delta t(3\omega_L)$
as well as the real amplitudes, $\chi^{(3)}_{1122}(\omega_L)$ and $\chi^{(3)}_{1122}(3\omega_L)$:
\begin{align}
  {p}^{(3)}_{1122}(t; \omega_L) =& \chi^{(3)}_{1122}(\omega_L) f^{(3)}[t - \delta{t}_{1122}(\omega_L); \omega_L] \label{eq:p3_delay1}\\
  {p}^{(3)}_{1122}(t; 3\omega_L) =& \chi^{(3)}_{1122}(3\omega_L) f^{(3)}[t - \delta{t}_{1122}(3\omega_L); 3\omega_L] \label{eq:p3_delay2}.
  \;
\end{align}
In the fitting procedure, we did not include the influence of the oscillating current after the pulse ends.
We show the obtained results employing LDA and TBmBJ in Table~\ref{tab:si}, including a comparison with measurements
and other calculations.

\begin{table*}[tb]
  \begin{tabular}{lccccccc}
    \hline
    &  & TDDFT &  & Previous Work &  &  & Experiments\\
    &  & LDA & TBmBJ & Semi ab-initio TB \cite{moss1991semi} & TDLDA & Opt. Pol. Func. \cite{gruning2016dielectrics} & \\ \hline
    \noalign{\smallskip}\noalign{\smallskip}
    $\chi^{(1)}$ &  & 15.2 & 12.4 &  &  &  & $12.8^\dagger$ Ref.\cite{ghosh1998handbook}\\
    \noalign{\smallskip}\noalign{\smallskip}
    $\chi^{(3)}_{1111}(\omega)$ [$\mathrm{m}^2/\mathrm{V}^2$] &  & $2.2\times10^{-18}$ & $8.6\times10^{-19}$ &  &  & $2.5\times10^{-18}$ & $1.4\times10^{-19\ast}$ Ref.\cite{bristow2007two}\\
     &  &  &  &  &  &  & $1.1\times10^{-19}$ Ref.\cite{boyd2003nonlinear}\\
    \noalign{\smallskip}\noalign{\smallskip}
    $\delta{t}^{(3)}_{1111}(\omega)$ [fs] &  & 0.6 & 0.2 &  &  &  & \\
    \noalign{\smallskip}\noalign{\smallskip}
    $\chi^{(3)}_{1111}(3\omega)$ [$\mathrm{m}^2/\mathrm{V}^2$] &  & $1.3\times10^{-18}$ & $1.4\times10^{-18}$ & $5.2\times10^{-19}$ & $1.1 \times 10^{-18}$ Ref.\cite{gruning2016dielectrics} &  & $2.8\times10^{-18}$ Ref.\cite{boyd2003nonlinear}\\
     &  &  &  &  & $3.6 \times 10^{-20\ast\ast}$ Ref.\cite{goncharov2013nonlinear} &  & \\
    \noalign{\smallskip}\noalign{\smallskip}
    $\delta{t}^{(3)}_{1111}(3\omega)$ [fs] &  & 0.7 & 0.6 &  &  &  & \\
    \noalign{\smallskip}\noalign{\smallskip}
    $3\chi^{(3)}_{1122}(\omega)$ [$\mathrm{m}^2/\mathrm{V}^2$] &  & $3.0\times10^{-18}$ & $1.2\times10^{-18}$ &  &  & $3.8\times10^{-18}$ & \\
    \noalign{\smallskip}\noalign{\smallskip}
    $\delta{t}^{(3)}_{1122}(\omega)$ [fs] &  & 0.6 & 0.2 &  &  &  & \\
    \noalign{\smallskip}\noalign{\smallskip}
    $3\chi^{(3)}_{1122}(3\omega)$ [$\mathrm{m}^2/\mathrm{V}^2$] &  & $2.3\times10^{-18}$ & $2.2\times10^{-18}$ & $5.7\times10^{-19}$ & $1.8 \times 10^{-18}$ Ref.\cite{gruning2016dielectrics} &  & \\
     &  &  &  &  & $5.2 \times 10^{-20\ast\ast}$ Ref.\cite{goncharov2013nonlinear} &  & \\
    \noalign{\smallskip}\noalign{\smallskip}
    $\delta{t}^{(3)}_{1122}(3\omega)$ [fs] &  & 0.7 & 0.2 &  &  &  & \\
    \noalign{\smallskip}\noalign{\smallskip}
    $|\sigma|$ &  & 0.7 & 0.6 & 0.2 & 0.6 Ref.\cite{gruning2016dielectrics} & 0.6 & $0.69^\dagger$, $0.85^{\dagger\dagger}$ Ref.\cite{moss1989dispersion}\\
    \noalign{\smallskip}\noalign{\smallskip}
    $\phi$ [deg.] &  & $6^{\circ}$ & $6^{\circ}$ & $11^{\circ}$ & $7^{\circ}$ Ref.\cite{gruning2016dielectrics} & $7^{\circ}$ Ref.\cite{gruning2016dielectrics} & $7^{\circ\dagger\dagger}$ Ref.\cite{moss1989dispersion}\\
     &  &  &  &  &  &  & $ 11^{\circ}$ Ref.\cite{moss1991semi}\\
    \noalign{\smallskip}\noalign{\smallskip}
    $E_G$ [eV] &  & 2.5 & 3.1 &  &  &  & 2.4 Ref.\cite{philipp1960optical} \\ \hline
  \end{tabular}
  \label{tab:c}
  \caption{
  Calculated linear and nonlinear susceptibilities of $\mathrm{Si}$.
  ${}^\dagger$ measurement at $\omega=1.6$~eV .
  ${}^\ast$ measurement at $\omega = 1.24$~eV .
  ${}^{\ast\ast}$ calculation at $\omega = 1.7$~eV .
  ${}^{\dagger\dagger}$ $\omega=1.60$~eV.
  ${}^{\dagger\dagger}$ $\omega=1.51$~eV.}
  \label{tab:si}
\end{table*}

In our calculations, linear susceptibility using the TBmBJ is close to the measured value, while the result using LDA is much larger.
For the third-order nonlinear susceptibilities, we find the values by LDA is about a factor 3-4 larger than the values by TBmBJ for $\chi^{(3)}(\omega_L)$, while there is not so much difference for $\chi^{(3)}(3\omega_L)$.
The time delay at $\omega \sim \omega_L$ region is much smaller in the TBmBJ calculation, indicating that the real excitation
is much less for the case of TBmBJ than the case of LDA.

Third-order nonlinear susceptibilities related to third-order harmonic generations have been traditionally discussed
using the complex quantities: $A=\chi^{(3)}_{1111}$, $B=3\chi^{(3)}_{1122}$, $\sigma=(A-B)/B$, and the angle $\phi(A/B)$.
We can construct the complex susceptibilities $\chi^{(3)}_{1111}(3\omega_L)$
from the magnitude $|\chi^{(3)}_{1111}(3\omega_L)|$ and the time delay $\delta t_{1111}(3\omega_L)$ as
$\chi^{(3)}_{1111}(3\omega_L) = | \chi^{(3)}_{1111}(3\omega_L)| e^{3i\omega_L \delta t_{1111}(3\omega_L)}$
and similar relations for other quantities.

Since values are rather scattered in both measurements and other theories, it is not simple to make a definite conclusion.
Apparently, more efforts are required. 
Our results are not so different from the TDLDA calculation by Gr\"uning et.al. \cite{gruning2016dielectrics}.
We find more than order of magnitude difference between our results and results by Goncharov \cite{goncharov2013nonlinear}, although both calculations use similar numerical method.

A direct comparison between two calculations is not simple because the induced polarization is treated differently.
In Ref. 26, the induced polarization is included in the applied electric field which we call the longitudinal geometry\cite{yabana2012time}, while we use the transverse geometry in which the induced polarization is not included. Although the final results should be the same in two schemes, quantities that appear in the intermediate steps are very different.
We also note that the magnitude of the electric field used to extract the nonlinear polarization components is different: In Ref.\cite{goncharov2013nonlinear}, it is mentioned that the electric field of 3 eV/A is used in the longitudinal geometry. This corresponds to the effective electric field of $\sim 3 \times 10^{-1}~\mathrm{eV/\AA}$ in the transverse geometry that we adopted.
In our calculation, as shown in Table I, we employ much weaker electric fields of $4 \times 10^{-2}~\mathrm{eV/\AA}$ or smaller.
In our experience, use of the electric fields larger than this value causes inaccuracy in the final results because of mixtures of components fifth order and higher.

\begin{figure*}
  \includegraphics[width=0.75\linewidth]{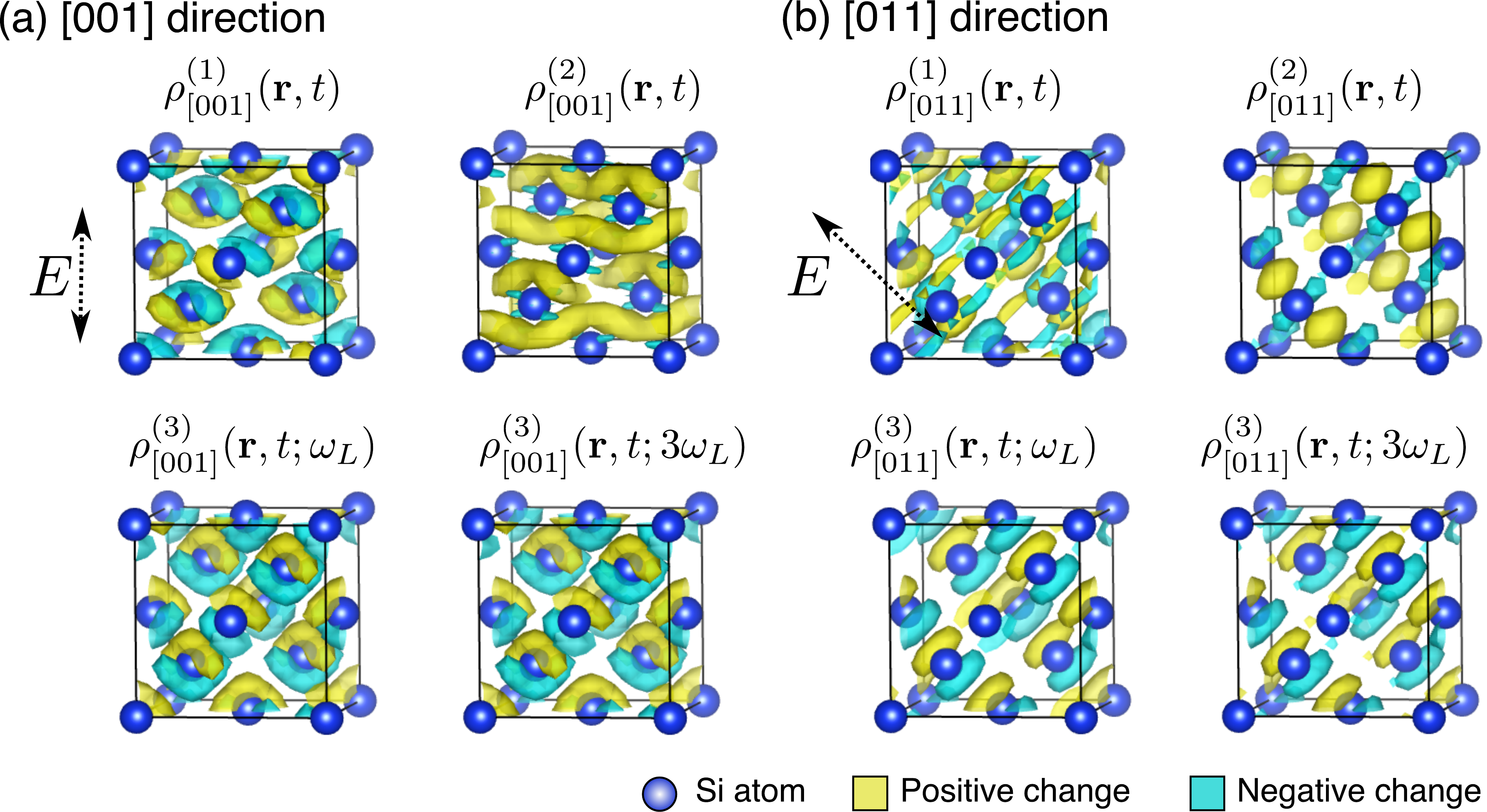}
  \caption{
  Electron density change of $\mathrm{Si}$ for two different directions of applied electric
  fields are decomposed into power series. For the third-order component, frequency
  filtering is applied.
  }
  \label{fig:6}       
\end{figure*}

In Fig.~\ref{fig:6}, we show decompositions of the electron density change for silicon
in the calculations of two directions, $[001]$ and $[011]$.
Linear, second-order, and third-order components are shown at a time $t \sim 10.6$ fs
where the electric field is maximum.
Surfaces of equal density changes are displayed in the unit cell volume.
Yellow surfaces indicate positive, and blue surfaces indicate negative changes.
For the third-order density changes, we achieve a frequency filtering as well as the
power series expansion with respect to the field amplitude.

Although linear optical response of silicon is isotropic and is characterized by a
single scalar dielectric function, the linear density change looks different dependent
on the direction of the applied field. We see in $\rho^{(1)}_{[011]}$, in which the electric
field is applied along the cubic axis, density oscillation along the bond direction is seen.
Nodal planes are seen perpendicular to the bond direction.
In $\rho^{(1)}_{[011]}$, in which the electric field is parallel to a part of the bonds,
the density oscillation is seen in the bonds which are parallel to the applied field.
Second-order density changes show a markedly different behavior dependent on the
direction of the electric field. In $\rho^{(2)}_{[001]}$, increase of electron density
connecting bonds is seen, while in $\rho^{(2)}_{[011]}$, movement of electrons between
orbitals perpendicular and parallel to the field is seen.
In the third-order density change, we first observe that the electron density change does
not differ much if we decompose it into the frequency components.
The electron density change looks more or less similar to the case of linear density
change.

\section{Summary}
\label{sec:summary}

Nonlinear polarization is a fundamental quantity that characterizes the interaction between
intense light and solids. It is important in both fundamental sciences and engineering applications.
Recently, measurements of nonlinear polarization in time domain become feasible by virtue of
the development of ultrashort laser technologies including attosecond metrologies.
It brings an opportunity to explore electron dynamics in real time and is also expected to provide
novel concepts of future signal processing using optical pulses. Searches for materials of high
nonlinear susceptibilities is also an important task.

In this work, we develop a theoretical and computational framework to explore nonlinear polarizations
in time domain induced by an ultrashort laser pulse based on first-principles time-dependent
density functional theory. We propose a numerical method to extract linear and nonlinear polarization
components, $\mathbf{p}^{(n)}(t) \quad (n=1~3)$, from solutions of the time-dependent Kohn-Sham
equation with the applied electric fields, $E(t) = E_i f(t)$, of a few different amplitudes, $E_i$.
The method is tested in three typical dielectrics, $\alpha$-$\mathrm{SiO}_2$, diamond, and silicon.
It has been shown that the method works accurately and reliably to extract the third-order
nonlinear polarization components if one uses electric fields with sufficiently small amplitudes.

The extracted nonlinear polarization components show characteristic features depending on
the optical bandgap energies of the materials and the frequency of the applied pulses, $\omega_L$.
In $\alpha$-SiO$_2$ that has a large optical gap, the third-order nonlinear polarization is essentially
proportional to the field amplitude cubed. For diamond for which the optical gap energy is close to
$3\omega_L$, nonlinear polarization shows complex time profile reflecting the frequency dependence
of the response. The complexity further increases for silicon in which real electronic excitations take
place by the third-order nonlinear process.

By fitting the extracted nonlinear polarization using the time profile of the applied field, we extract
the nonlinear susceptibilities and the time delay. The extracted coefficients are compared with
measurements and previous theoretical calculations.

We also show that it is possible to decompose the electron density change from the ground state
into linear and nonlinear components. The method is expected to be useful to get an intuitive
picture for the electron dynamics in dielectrics and to understand the origin of the nonlinear
susceptibilities.

\begin{acknowledgments}
  We acknowledge the supports by MEXT as a priority issue theme 7 to be 
  tackled by using
  Post-K Computer, and JST-CREST under grant number JP-MJCR16N5, and by
  JSPS KAKENHI Grant Numbers 15H03674.
  Calculations are carried out at Oakforest-PACS at JCAHPC through the 
  Multidisciplinary
  Cooperative Research Program in CCS, University of Tsukuba, and through 
  the HPCI System
  Research Project (Project ID: hp180088).
\end{acknowledgments}

\bibliography{refs}

\end{document}